\documentclass[aps,prl,twocolumn,superscriptaddress]{revtex4}

\usepackage{graphicx}
\usepackage[german,english]{babel}
\newcommand{\diff}{\ensuremath{\text{d}}}
\newcommand{\sgn}{\ensuremath{\text{sgn}}}

\newcommand{\Tr}{\,{\rm{Tr}\,}}

\renewcommand{\Im}{{\rm Im}}

\newcommand{\beqa}{\begin{eqnarray}}
\newcommand{\eeqa}{\end{eqnarray}}

\begin{document}

\title{Local impurity effects in superconducting graphene}

\author{T. O. Wehling}
\affiliation{I. Institut f{\"u}r Theoretische Physik, Universit{\"a}t Hamburg, Jungiusstra{\ss}e 9, D-20355 Hamburg, Germany}
\author{H. P. Dahal}
\affiliation{Theoretical Division, Los Alamos National Laboratory, Los Alamos, New Mexico 87545,USA}

\author{A. I. Lichtenstein}
\affiliation{I. Institut f{\"u}r Theoretische Physik, Universit{\"a}t Hamburg, Jungiusstra{\ss}e 9, D-20355 Hamburg, Germany}

\author{A. V.  Balatsky}
\email[]{avb@lanl.gov, http://theory.lanl.gov}
\affiliation{Theoretical Division, Los Alamos National Laboratory, Los Alamos, New Mexico 87545,USA}
\affiliation{Center for Integrated
Nanotechnologies, Los Alamos National Laboratory, Los Alamos, New
Mexico 87545,USA}


\date{\today}

\begin{abstract}
We study the effect of impurities in superconducting graphene and
discuss their influence on the local electronic properties. In
particular, we consider the case of magnetic and non-magnetic
impurities being either strongly localized or acting as a potential
averaged over one unit cell. The spin dependent local density of
states is calculated and possibilities for visualizing impurities by
means of scanning tunneling experiments is pointed out. A
possibility of identifying magnetic scatters even by non
spin-polarized scanning tunneling spectroscopy is explained.
\end{abstract}


\maketitle

Graphene - one monolayer of carbon atoms arranged in a honeycomb
lattice - can be considered as one of the most promising materials
for post-silicon electronics \cite{Novoselov_science2004}. Impurity states often are the basis of
modern electronic devices. Because of
its peculiar electronic structure graphene provides a model system for
studying relativistic quantum physics in condensed matter
\cite{Geim2005, Zhang2005}. A careful investigation of the
impurity states in graphene also allows one to address fundamental
questions of relativistic scattering
\cite{bena:2005,peres:125411,Loktev-2006-,cheianov:226801,imp_loc-2007-}.
Thus, far the  central component to these studies is graphene's
linearly vanishing density of states (DOS) near the Dirac point: It
allows for virtual bound states (VBS) in this region, i.e.
resonances in the DOS, which may be arbitrarily sharp in the
vicinity of the Dirac point. The broad similarities of these
impurity states with states in d-wave superconductors have been used
to discuss common aspects of impurity scattering in superconductors
and in graphene
\cite{imp_loc-2007-,balatsky:373,Loktev-2006-,peres:125411,pereira2006}.

Recently, it has been demonstrated that it is possible to induce
superconductivity in graphene via a proximity effect
\cite{heersche:2007}. Prior analysis of the ballistic
superconducting transport \cite{titov:041401} revealed an
interesting suppression of the critical current near the Dirac
point. It is therefore simply a matter of time before the defects
and impurity induced states in superconducting graphene will be
addressed locally. The case of superconductivity in graphene where
opposite valleys are nontrivially involved is also an explicit
example of supercondcutivity in vallyetronics \cite{Rycerz07}. The
observation of the proximity effect in graphene raised fundamentally
new questions about impurity effects in this material in presence of
superconducting correlations: i) Is there a possibility of intragap
bound states and ii) what is the impact of gap opening on the
Friedel oscillations in the continuum?

In this letter we show that magnetic impurities do produce impurity
bound states {\em inside the superconducting gap}. These bound
states \textit{always} coexist with the formerly studied VBS in the
continuum. Thus the predicted impurity states are similar to the
magnetic impurity induced states, so called Yu Lu-Shiba-Rusinov
states, in s-wave superconductors \cite{AliYazdani1997,balatsky:373}.
Due to its 2 dimensionality graphene is well suited for Scanning
Tunneling Microscopy (STM) investigations and first experiments on
normal state graphene already indicate the importance of impurity
effects in this context \cite{ClaireBerger05262006,mallet-2007}.
Therefore, we elucidate the real space shape of these impurity
states, which will be directly observable in STM experiments.

While graphene intrinsically is not superconducting, the Ti/Al
bilayer contacts placed on the graphene sheet induce a measured
supercurrent \cite{heersche:2007}. No spectral gap in the samples
has been measured to date. We argue that the residual
electron-electron interaction in the graphene will produce a {\em
gap} in the spectrum. This gap will be proportional to the
interaction strength and it remains to be seen how large this gap
can be in the graphene. Electron spectroscopy such as STM and/or
planar tunneling into graphene in proximity to superconducting leads
would be able to reveal the spectroscopic gap. We will treat
superconducting gap $\Delta$ below as a phenomenological parameter
that needs to be determined separately.

Low energy electronic excitations in graphene can be described by
two species of Dirac fermions located around two nodal points
$K^\pm$ in the Brillouin zone with the speed of light being replaced
by the Fermi velocity $v_{\rm f}$ and the corresponding Hamiltonian
$H_{K^\pm}=v_{\rm f}\hbar(k_1\sigma_1\mp k_2\sigma_2)$. $\sigma_i$,
$i=1,2,3$, are Pauli matrices acting on the sublattice degrees of
freedom and $\sigma_0$ is the identity matrix. To understand
impurities in superconducting graphene, we use the Nambu formalism
including both valleys:
\[\hat{H}=-i\hbar
v_f\int\diff^2x\hat{\Psi}^\dagger(x)(\partial_1\sigma_1\otimes\tau_0-\partial_2\sigma_2\otimes\tau_3)\otimes\Lambda_0\hat{\Psi}(x)\]
with $\hat{\Psi}(x)^\dagger=(\Psi_{\downarrow
K^+}^\dagger(x),\Psi_{\downarrow K^-}^\dagger(x),\Psi_{\uparrow
K^-}(x),\Psi_{\uparrow K^+}(x))$ and $\Psi_{\uparrow\downarrow
K^\pm}(x)$ being field operators of electrons with a spin
$\uparrow\downarrow$ and belonging to a valley $K^\pm$. $\tau_i$ and
$\Lambda_i$ with $i=1,2,3$ are Pauli matrices but acting on the
valley and Nambu space, respectively. $\tau_0$ and $\Lambda_0$ are
the corresponding identity matrices. In contact with a
superconductor, the proximity effect imposes a finite pairing
potential $\Delta\sigma_3\otimes\tau_0\otimes\Lambda_1$ to the
graphene sheet and results in electron dynamics being described by
the Dirac-Bogoliubov-de Gennes (DBdG) Hamiltonian
\cite{beenakker:067007}:
\begin{equation}
\label{eqn:Hamiltonian-SC}
H=-i\hbar v_f(\partial_1\sigma_1\otimes\tau_0-\partial_2\sigma_2\otimes\tau_3)\otimes\Lambda_0+\Delta\sigma_3\otimes\tau_0\otimes\Lambda_1.
\end{equation}
\begin{figure}
\begin{minipage}{.40\linewidth}
\centering
\includegraphics[width=.99\linewidth]{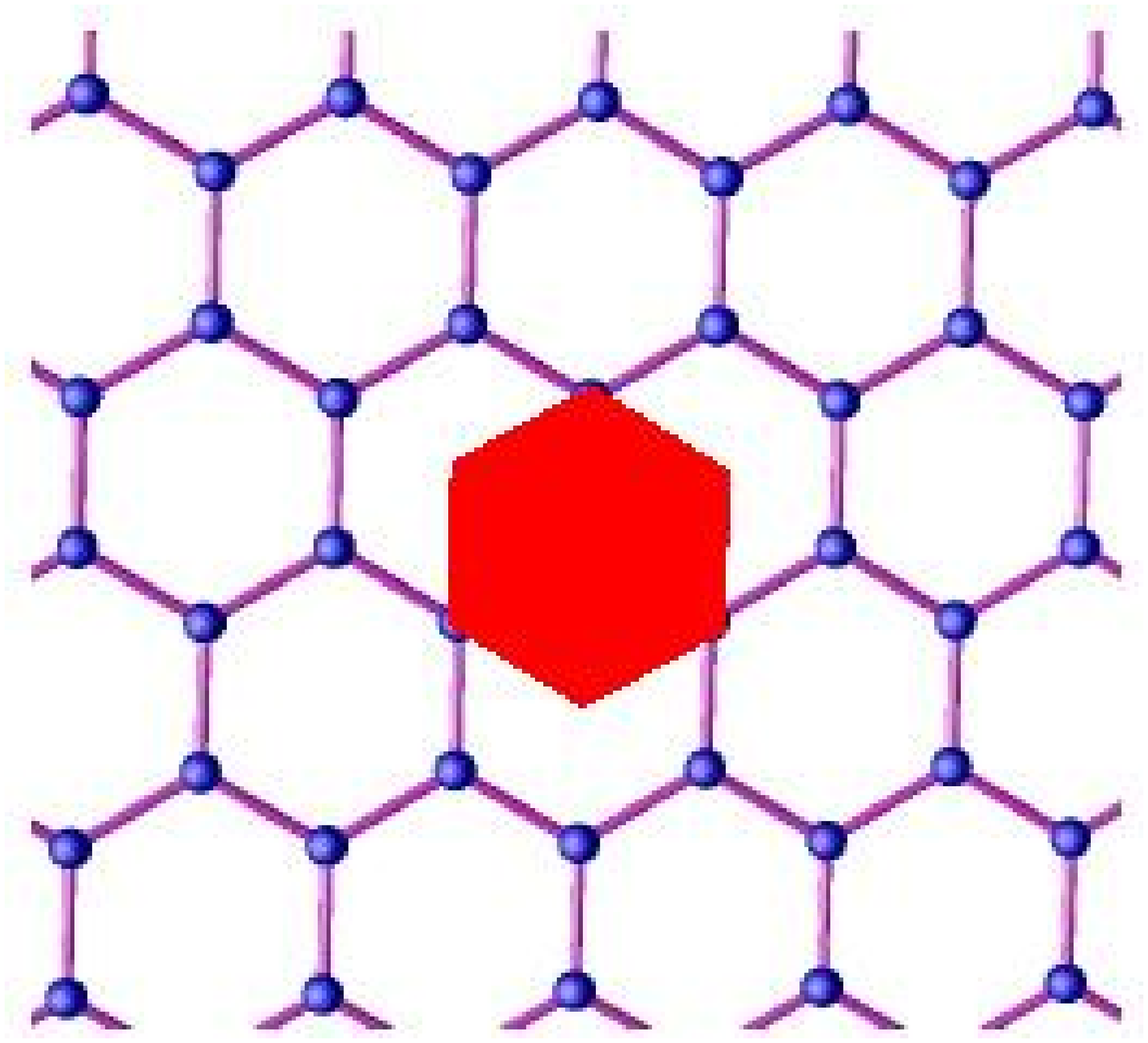}
\end{minipage}\hfill\begin{minipage}{.40\linewidth}
\centering
\includegraphics[width=.99\linewidth]{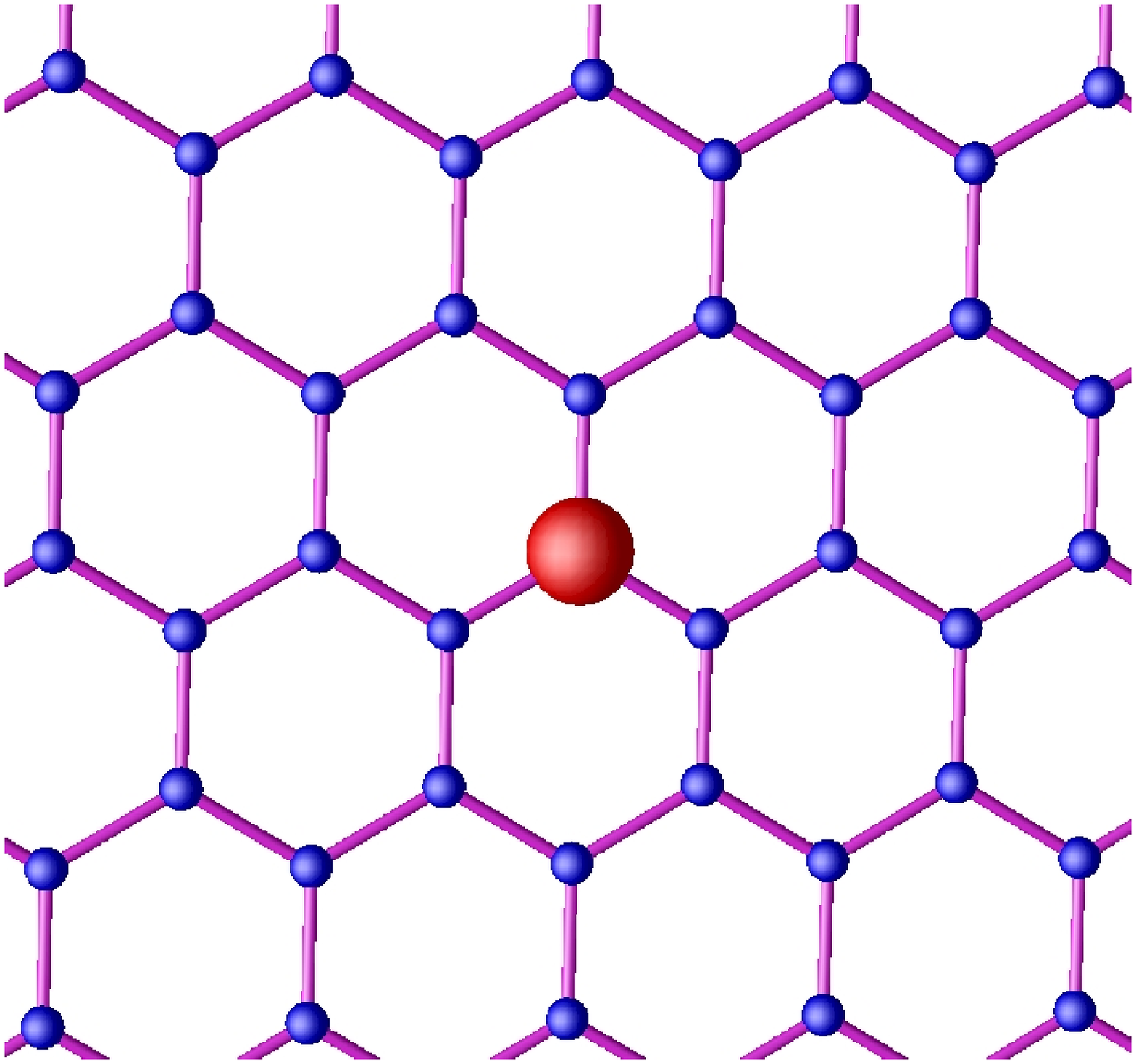}
\end{minipage}
\caption{\label{fig:ImpTypes}(Color online) Among the various local
impurities we discuss two limiting cases. The scalar impurity
(left), $V_s$, corresponds to a uniform potential averaged over one
unit cell, whereas the on-site impurity (right), $V_o$, acts on one
sublattice only.}
\end{figure}
To elucidate the effect of different impurities, we discuss both, a
homogeneous potential acting within one unit cell $V_{\rm s}$
(referred to as scalar impurity, Fig. \ref{fig:ImpTypes}.a) as well
as a strongly localized impurity $V_o$ (referred to as on-site
impurity, Fig. \ref{fig:ImpTypes}.b) acting only at sublattice $A$
and giving rise to intervalley scattering.

Starting from impurity operators in the tight-binding form of e.g.
Ref. \cite{imp_loc-2007-} and using the conventions of Eqn.
(\ref{eqn:Hamiltonian-SC}) we obtain the following explicit
expressions for the impurity potentials in the adopted matrix
notation: $V_{\rm
s}=V_0\sigma_0\otimes\tau_0\otimes\Lambda_3+V_1\sigma_0\otimes\tau_0\otimes\Lambda_0$
and
$V_o=V_0(\sigma_3+\sigma_0)\otimes(\tau_0+\tau_1)\otimes\Lambda_3+V_1(\sigma_3+\sigma_0)\otimes(\tau_0+\tau_1)\otimes\Lambda_0$
\cite{technote07}. In both cases $V_0$ and $V_1$ describe the
electrostatic and magnetic contribution to the impurity potential,
respectively.

The effects of these impurities on the local electronic properties
of the superconducting graphene sheets is contained in the local
density of states (LDOS), which we calculate using the T-matrix
approach \cite{balatsky:373}: In operator form, the full Green's
function $G(\omega)$ in presence of the impurity is obtained from
the unperturbed Green's function $G^0(\omega)$ via
$G(\omega)=G^0(\omega)+G^0(\omega)T(\omega)G^0(\omega)$ with
$T(\omega)=V_{s(o)}(1-G^0(\omega)V_{s(o)})^{-1}$.

Dealing with local impurities, it is convenient to adopt the
position space representation. Therefore, the free $x$-dependent
Green's function $\hat{G}^0(x,\omega)$ in polar coordinates,
$x=x(r,\phi)$, is obtained from its momentum space counterpart
$\hat{G}^0(p,\omega)=(\omega-H)^{-1}=\frac{(\omega\sigma_0\otimes\tau_0+v_{\rm
f}[p_1\sigma_1\otimes\tau_0-p_2\sigma_2\otimes\tau_3])\otimes\Lambda_0+\Delta\sigma_3\otimes\tau_0\otimes\Lambda_1}{\omega^2-v_{\rm
f}^2p^2-\Delta^2}$
by Fourier transformation \begin{widetext}\begin{equation}
\hat{G}^0(x,\omega) = \int\frac{\diff^2 p}{\Omega_B}
\,\hat{G}^0(p,\omega)e^{ipx} =
g_0(r,\omega)(\omega\sigma_0\otimes\tau_0\otimes\Lambda_0+\Delta\sigma_3\otimes\tau_0\otimes\Lambda_1)+
g_1(r,\omega)([\cos\phi\,\sigma_1\otimes\tau_0+\sin\phi\,\sigma_2\otimes\tau_3]\otimes\Lambda_0)
\label{eqn:FT_G_RealSp}
\end{equation}
\end{widetext}
with $g_0(r,\omega)=v_{\rm f}^2\int_0^{p_c}\diff
p\,pJ_0(pr)(W^2(\omega^2-\Delta^2-v_{\rm f}^2p^2))^{-1}$ and
$g_1(r,\omega)=iv_{\rm f}^3\int_0^{p_c}\diff
p\,p^2J_1(pr)(W^2(\omega^2-\Delta^2-v_{\rm f}^2p^2))^{-1}$, where we
expressed the Brillouin zone volume $\Omega_B=2\pi W^2/v_{\rm f}^2$
in terms of the bandwidth $W$. The Green's function at $x=0$
determines the LDOS of the free system and it occurs in the
T-matrix:
$\hat{G}^0(0,\omega+i\delta)=M(\omega)(\omega\sigma_0\otimes\tau_0\otimes\Lambda_0+\Delta\sigma_3\otimes\tau_0\otimes\Lambda_1).$
Here is $M(\omega)=M'(\omega)+iM''(\omega)$ with
$M'(\omega)=\frac{1}{2W^2}\ln\left|\frac{\Delta^2-\omega^2}{W^2+\Delta^2-\omega^2}\right|$
and $M''(\omega)= -\frac{\pi\sgn(\omega)}{2W^2}$ for
$\Delta^2<\omega^2<\Delta^2+W^2$ and $M''(\omega)=0$ else. One sees, that the corresponding LDOS 
vanishes within the superconducting gap ($\omega^2<\Delta^2$) and is
given by $N_0(\omega)=\frac{4|\omega|}{W^2}$ outside the gap.

\begin{figure}[t]
\includegraphics[width=.80\linewidth]{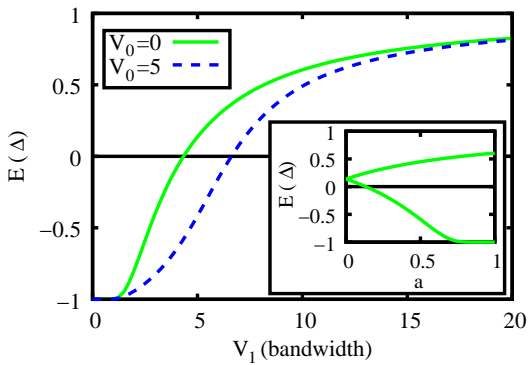}
\caption{\label{fig:EimpU}(Color online) Energy of the impurity
resonance for the scalar impurity as a function of the magnetic
impurity potential $V_1$ for different electrostatic potentials
$V_0$. The gap-parameter is $\Delta=W/10$. The lower right inset shows
the splitting of the impurity state due to intervalley scattering.
We model intervalley scattering as $V = V_1(\sigma_0 \otimes \tau_0
\otimes \Lambda_0 + a \sigma_0 \otimes \tau_1 \otimes \Lambda_0)$
with the strength of the intervalley scattering parametrized by $a$
and $V_1=5W$. One split state is shifted to the gap edge, the
other state remains an intragap state.}
\end{figure}
In general, impurity resonances occur when the $T$ matrix becomes (almost) singular, i.e. $\det(1-G^0(0,\omega)V)=0$. 
For the scalar impurity this secular equation yields
$1-2M(\omega)\omega
V_1+M^2(\omega)(\omega^2-\Delta^2)(V_1^2-V_0^2)=0$ with solutions
that can be understood analytically in the following limiting cases:
Firstly consider a solely magnetic impurity, i.e. $V_0=0$, with
$V_1>0$. In the Born limit the solutions
$\omega_0=-\Delta\pm\delta\omega$ with
\begin{equation}
\delta\omega=\frac{W^2}{2\Delta}e^{-W^2/(\Delta V_1)}
\end{equation}
give rise to intragap bound and virtual bound states in the
continuum approaching the gap edge exponentially with decreasing
$V_1$. In the opposite limit of unitary scattering
$\omega_0=\pm\Delta-\delta\omega$ with
\begin{equation}\delta\omega=-\frac{2W^2}{V_1\ln\left(\frac{2\Delta}{V_1}\right)}\end{equation}
fulfills the secular equation, where the upper (lower) sign
corresponds to a intragap bound (continuum virtual bound) state.

The numerical solutions to energies of the intragap bound states are
shown in Fig. \ref{fig:EimpU}. It recovers the limiting cases
obtained analytically and demonstrates also the effect of an
electrostatic contribution $V_0$ to the impurity potential: In the
Born limit, the exponential dependence of $\delta\omega$ on the
magnetic potential strength $V_1$ is dominant and suppresses any
significant influence of $V_0$ on the impurity state energy. In the
$V_1\rightarrow\infty$ limit, $V_0$ leads to a renormalization of
the \textit{effective} magnetic potential strength $V_1\rightarrow
V_1(1-\frac{V_0^2}{V_1^2})$. As Fig. \ref{fig:EimpU} shows, the
effect of an additional electrostatic potential becomes most
pronounced in the intermediate region. There, the electrostatic
contribution reduces the effective magnetic potential strength most
significantly.


Having understood the energy of intra gap bound states due to scalar
impurities, we address now the strongly localized on-site
impurities and elucidate the effect of inter valley scattering. Due to valley degeneracy, the scalar impurity gives rise to doubly
degenerate intra gap bound states. This degeneracy is
lifted by intervalley scattering, see Fig.
(\ref{fig:EimpU}). The
secular equation corresponding to the on-site impurity, $1-8M(\omega)\omega
V_1+16M^2(\omega)(\omega^2-\Delta^2)(V_1^2-V_0^2)=0$, reduces to that
of an on-site impurity with the replacement $V_{0,1}\rightarrow
4V_{0,1}$. Besides the lifting of the valley degeneracy, additional
intervalley scattering results also in a renormalization of the
effective impurity strength.

With the real-space Green's function
$G(x,x',\omega)=G^0(x-x',\omega)+G^0(x,\omega)T(\omega)G^0(-x',\omega)$
one obtains the local density of states
$N(x,\omega)=N_0(\omega)+\delta N(x,\omega)=-\frac{1}{\pi}\Im
G(x,x,\omega)$ in presence of an impurity. This LDOS is a matrix
corresponding to the matrix structure of the Green's function. It
accounts for the contributions from the different sublattices,
valleys and the Nambu space. According to the convention in Eqn.
(\ref{eqn:Hamiltonian-SC}), the spin-up excitations are hole
excitations yielding for each spin component the LDOS
$N_{\downarrow\uparrow}(x,\omega)=\Tr\frac{\Lambda_0\pm\Lambda_3}{2}N(x,\pm\omega)$,
where the trace involves either the spin-down or up part of the
Nambu space. In the case of the scalar impurity, this yields
explicitly the following corrections to the unperturbed LDOS in the
continuum
\begin{widetext}
\begin{equation}
\label{eqn:LDOS_scalar}
\delta N_{\downarrow\uparrow}(r,\pm\omega)=-\frac{4}{\pi}\Im\frac{a_{\downarrow\uparrow} g_0^2(r,\omega)+b_{\downarrow\uparrow} g_1^2(r,\omega)}{1-2M(\omega)\omega V_1+M^2(\omega)(\omega^2-\Delta^2)(V_1^2-V_0^2)}
\end{equation}
with $a_{\downarrow\uparrow}=(\omega^2-\Delta^2)[\pm
V_0+M(\omega)\omega(V_0^2-V_1^2)]+(\omega^2+\Delta^2)V_1$ and
$b_{\downarrow\uparrow}=(\pm V_0 + V_1) + M(\omega)\omega (V_0^2 -
V_1^2)$.
\end{widetext}
By replacing $M(\omega)\rightarrow 4M(\omega)$ in
these formula, one obtains the case of the strongly localized
on-site impurity.
In STM experiments, graphene's lattice structure will give rise to a
triangular modulation of the impurity states. This is neglected
here, as similar effects in normal state graphene have been
discussed in Ref. \cite{imp_loc-2007-}.

Due to Eqn. (\ref{eqn:LDOS_scalar}), the asymptotic decay of the
intra gap bound states at large distances from the impurity is
governed by $g_0^2(r,\omega_0)$ and $g_1^2(r,\omega_0)$. Neglecting
high-energy cut-off related oscillations at this length scale, one
may extend the momentum space integrals in Eqn.
(\ref{eqn:FT_G_RealSp}) to infinity. This yields modified Bessel
functions, i.e.
$g_0(r,\omega_0)=-\frac{1}{W^2}\rm{K}(0,r\sqrt{\Delta^2 -
\omega_0^2}/v_{\rm f})$ and $g_1(r,\omega_0)=-\frac{i\sqrt{\Delta^2
- \omega_0^2}}{W^2}\rm{K}(1,r\sqrt{\Delta^2 - \omega_0^2}/v_{\rm
f})$. Therefore the wavefunctions of the impurity states decay as
\begin{equation}
\delta N_{\downarrow\uparrow}(r,\pm\omega_0)\propto r^{-1}e^{-2r\sqrt{\Delta^2 -\omega_0^2}/v_{\rm f}}.
\end{equation}
As Fig. \ref{fig:WF_scalar_onsite} (left) shows, impurity states in
the gap give rise to prominent features in future STM experiments:
\begin{figure}
\includegraphics[width=.95\linewidth]{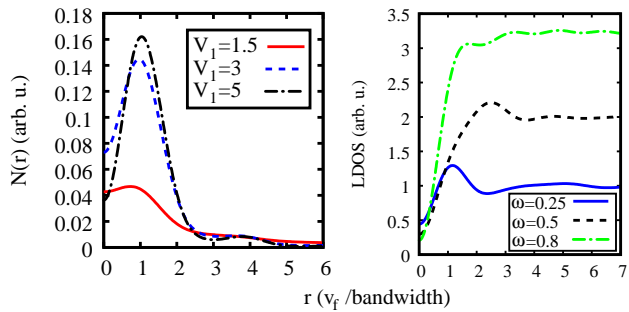}
\caption{\label{fig:WF_scalar_onsite}(Color online) Left panel:
Density $N$ of the intra-gap bound states as a function of the
distance $r$ from the impurity for purely magnetic scalar impurities
and different potentials $V_1$. The impurity strength is given in
units of the bandwidth $W$. Right panel: Friedel oscillations in the
local density of states (LDOS) around a scalar impurity at $r=0$
with $V_0=0$ and $V_1=3W$. The different curves correspond to the
energies $\omega=0.8$, $0.5$ and $0.25W$. In both panels, the
gap-parameter is $\Delta=W/10$.}
\end{figure}
The density of the impurity state at the impurity site at $r=0$ as
well as the maximum of the density are strongly sensitive to the
particular type of impurity. In general, impurity states with
energies in the middle of the gap ($V_1=5W$ in Fig.
\ref{fig:WF_scalar_onsite} (left)) give rise to the sharpest maxima
in the $r$-dependent LDOS. The ratio of the maximum density to the
density at the impurity site increases with the potential strength
$V_1$.

The ringstructure corresponding to these impurity states in STM
images may give a powerful experimental tool for identifying
particular impurities present in superconducting graphene. This is
in contrast to the normal state graphene, where weak impurities do
not give rise to resonances near the Dirac point and will therefore
hardly be apparent in scanning tunneling spectroscopy (STS)\cite{imp_loc-2007-}.

In the continuum, Eqn. (\ref{eqn:LDOS_scalar}) encodes the real
space shape of VBS and Friedel oscillations around the impurities.
As Fig. \ref{fig:WF_scalar_onsite} (right) shows exemplarily for a
scalar impurity, the wavelength $\lambda$ of these oscillations is
in any case determined by the energy $\omega$ and the gap $\Delta$:
$\lambda=\pi v_f/\sqrt{\omega^2-\Delta^2}$. Besides these
oscillations giving rise to standing wave patterns in future STM
experiments, certain resonances due to VBS will be an even more
prominent as well as impurity specific feature in these experiments:
The LDOS in Fig. \ref{fig:WF_scalar_onsite} (right) exhibits a
characteristic peak at $r\approx 1$ and $\omega\approx 0.25W$.

So far, we have discussed the r-dependent LDOS for different
impurities and at different energies corresponding to STM images at
fixed bias. The impurities will manifest themselves also in the
energy dependence of the LDOS at fixed position, which is accessible
by STS.
\begin{figure}
\includegraphics[width=.95\linewidth]{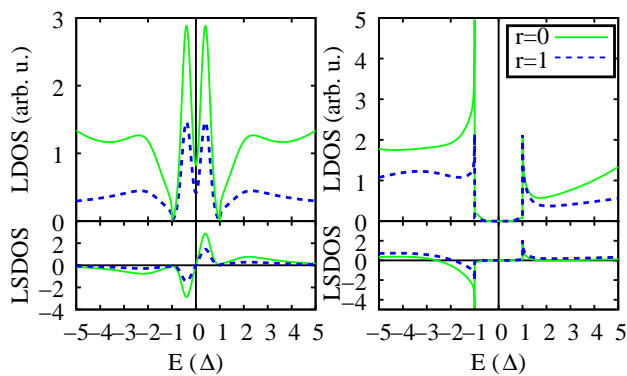}
\caption{\label{fig:E-LDOS-scalar-onsite}(Color online) The local
density of states (LDOS, upper panel) and the local spin density of
states (LSDOS, lower panel), $\delta N_{\uparrow}(r,\omega)-\delta
N_{\uparrow}(r,\omega)$, at $r=0$ and $r=1$ is shown for scalar of
impurities with different potentials: a purely magnetic impurity
with $V_0=0$ and $V_1=3W$ (left) as well as an impurity with
$V_0=2W$ and $V_1=1W$.}
\end{figure}
In Fig. \ref{fig:E-LDOS-scalar-onsite}, the LDOS near a purely
magnetic scalar impurity with $V_0=0$ and $V_1=3W$ is compared to an
impurity contributing an electrostatic potential $V_0=2W$ and
$V_1=1W$. The purely magnetic impurity, Fig.
\ref{fig:E-LDOS-scalar-onsite} (left), does not break particle hole
symmetry and yields therefore a fully symmetric LDOS and a fully
antisymmetric local spin density of states (LSDOS). This is in
contrast to the more general second impurity, Fig.
\ref{fig:E-LDOS-scalar-onsite} (right), where the LDOS and LSDOS are
not symmetric under  particle hole transformation. Therefore, the
degree of symmetry of the local spectra allows to estimate, whether
the impurity potential is magnetic or not - even in a non-spin
polarized scanning tunneling spectroscopy experiment.

In \textit{conclusion}, we argued that magnetic scattering will
produce impurity induced bound and virtual bound states in
superconducting graphene. These impurity states are similar to the
Yu Lu-Shiba-Rusinov states in s-wave superconductors
\cite{balatsky:373} and exhibit an intricate real space and
particle-hole dependent structure. We discussed the energy
dependence of these states as a function of the potential parameters
and pointed out characteristic oscillation pattern as well as decay
properties in real space. This spectroscopic and topographic
information can be obtained by STM \cite{AliYazdani1997}. We showed
that each impurity generates a specific signature in the real space
LDOS and provided a guideline for identifying different impurities
in future experiments. Since the Cooper pairs have zero momentum we
find superconducting state in graphene to be a nontrivial example of
valleytronics where valley quantum numbers are important
\cite{Rycerz07}.

The authors thank E. Andrei, C. Beenakker,  A. H. Castro Neto, H.
Fukuyama, A. Geim, P. J. Hirschfeld, M. I. Katsnelson, A. F. Morpurgo, I. Vekhter and
J. X. Zhu for useful discussions. This work was supported by US DOE
at Los Alamos and SFB 668. T.O.W. is grateful to LANL and the T11 group for hospitality during the visit,
when the ideas presented in this work were conceived.
\bibliography{graphene_s}

\begin{thebibliography}{18}
\expandafter\ifx\csname natexlab\endcsname\relax\def\natexlab#1{#1}\fi
\expandafter\ifx\csname bibnamefont\endcsname\relax
  \def\bibnamefont#1{#1}\fi
\expandafter\ifx\csname bibfnamefont\endcsname\relax
  \def\bibfnamefont#1{#1}\fi
\expandafter\ifx\csname citenamefont\endcsname\relax
  \def\citenamefont#1{#1}\fi
\expandafter\ifx\csname url\endcsname\relax
  \def\url#1{\texttt{#1}}\fi
\expandafter\ifx\csname urlprefix\endcsname\relax\def\urlprefix{URL }\fi
\providecommand{\bibinfo}[2]{#2}
\providecommand{\eprint}[2][]{\url{#2}}

\bibitem[{\citenamefont{Novoselov et~al.}(2004)}]{Novoselov_science2004}
\bibinfo{author}{\bibfnamefont{K.~S.} \bibnamefont{Novoselov}}
  \bibnamefont{et~al.}, \bibinfo{journal}{Science}
  \textbf{\bibinfo{volume}{306}}, \bibinfo{pages}{666} (\bibinfo{year}{2004}).

\bibitem[{\citenamefont{Novoselov et~al.}(2005)}]{Geim2005}
\bibinfo{author}{\bibfnamefont{K.~S.} \bibnamefont{Novoselov}}
  \bibnamefont{et~al.}, \bibinfo{journal}{Nature}
  \textbf{\bibinfo{volume}{438}}, \bibinfo{pages}{197} (\bibinfo{year}{2005}).

\bibitem[{\citenamefont{Zhang et~al.}(2005)}]{Zhang2005}
\bibinfo{author}{\bibfnamefont{Y.}~\bibnamefont{Zhang}} \bibnamefont{et~al.},
  \bibinfo{journal}{Nature} \textbf{\bibinfo{volume}{438}},
  \bibinfo{pages}{201} (\bibinfo{year}{2005}).

\bibitem[{\citenamefont{Bena and Kivelson}(2005)}]{bena:2005}
\bibinfo{author}{\bibfnamefont{C.}~\bibnamefont{Bena}} \bibnamefont{and}
  \bibinfo{author}{\bibfnamefont{S.~A.} \bibnamefont{Kivelson}},
  \bibinfo{journal}{Phys. Rev. B} \textbf{\bibinfo{volume}{72}},
  \bibinfo{pages}{125432} (\bibinfo{year}{2005}).

\bibitem[{\citenamefont{Peres et~al.}(2006)}]{peres:125411}
\bibinfo{author}{\bibfnamefont{N.~M.~R.} \bibnamefont{Peres}}
  \bibnamefont{et~al.}, \bibinfo{journal}{Phys. Rev. B}
  \textbf{\bibinfo{volume}{73}}, \bibinfo{eid}{125411} (\bibinfo{year}{2006}).

\bibitem[{\citenamefont{Skrypnyk and Loktev}(2006)}]{Loktev-2006-}
\bibinfo{author}{\bibfnamefont{Y.~V.} \bibnamefont{Skrypnyk}} \bibnamefont{and}
  \bibinfo{author}{\bibfnamefont{V.~M.} \bibnamefont{Loktev}},
  \bibinfo{journal}{Phys. Rev. B} \textbf{\bibinfo{volume}{73}},
  \bibinfo{pages}{241402 (R)} (\bibinfo{year}{2006}).

\bibitem[{\citenamefont{Cheianov and Fal'ko}(2006)}]{cheianov:226801}
\bibinfo{author}{\bibfnamefont{V.~V.} \bibnamefont{Cheianov}} \bibnamefont{and}
  \bibinfo{author}{\bibfnamefont{V.~I.} \bibnamefont{Fal'ko}},
  \bibinfo{journal}{Phys. Rev. Lett.} \textbf{\bibinfo{volume}{97}},
  \bibinfo{eid}{226801} (\bibinfo{year}{2006}).

\bibitem[{\citenamefont{Wehling et~al.}(2007)}]{imp_loc-2007-}
\bibinfo{author}{\bibfnamefont{T.~O.} \bibnamefont{Wehling}}
  \bibnamefont{et~al.}, \bibinfo{journal}{Phys. Rev. B}
  \textbf{\bibinfo{volume}{75}}, \bibinfo{eid}{125425} (\bibinfo{year}{2007}).

\bibitem[{\citenamefont{Balatsky et~al.}(2006)}]{balatsky:373}
\bibinfo{author}{\bibfnamefont{A.~V.} \bibnamefont{Balatsky}}
  \bibnamefont{et~al.}, \bibinfo{journal}{Rev. Mod. Phys.}
  \textbf{\bibinfo{volume}{78}}, \bibinfo{eid}{373} (\bibinfo{year}{2006}).

\bibitem[{\citenamefont{Pereira et~al.}(2006)}]{pereira2006}
\bibinfo{author}{\bibfnamefont{V.~M.} \bibnamefont{Pereira}}
  \bibnamefont{et~al.}, \bibinfo{journal}{Phys. Rev. Lett.}
  \textbf{\bibinfo{volume}{96}}, \bibinfo{eid}{036801} (\bibinfo{year}{2006}).

\bibitem[{\citenamefont{Heersche et~al.}(2007)}]{heersche:2007}
\bibinfo{author}{\bibfnamefont{H.~B.} \bibnamefont{Heersche}}
  \bibnamefont{et~al.}, \bibinfo{journal}{Nature}
  \textbf{\bibinfo{volume}{446}}, \bibinfo{pages}{56} (\bibinfo{year}{2007}).

\bibitem[{\citenamefont{Titov and Beenakker}(2006)}]{titov:041401}
\bibinfo{author}{\bibfnamefont{M.}~\bibnamefont{Titov}} \bibnamefont{and}
  \bibinfo{author}{\bibfnamefont{C.~W.~J.} \bibnamefont{Beenakker}},
  \bibinfo{journal}{Phys. Rev. B} \textbf{\bibinfo{volume}{74}},
  \bibinfo{pages}{041401(R)} (\bibinfo{year}{2006}).

\bibitem[{\citenamefont{Rycerz et~al.}(2007)}]{Rycerz07}
\bibinfo{author}{\bibfnamefont{A.}~\bibnamefont{Rycerz}} \bibnamefont{et~al.},
  \bibinfo{journal}{Nature Phys.} \textbf{\bibinfo{volume}{3}},
  \bibinfo{pages}{172} (\bibinfo{year}{2007}).

\bibitem[{\citenamefont{Yazdani et~al.}(1997)}]{AliYazdani1997}
\bibinfo{author}{\bibfnamefont{A.}~\bibnamefont{Yazdani}} \bibnamefont{et~al.},
  \bibinfo{journal}{Science} \textbf{\bibinfo{volume}{275}},
  \bibinfo{pages}{1767} (\bibinfo{year}{1997}).

\bibitem[{\citenamefont{Berger et~al.}(2006)}]{ClaireBerger05262006}
\bibinfo{author}{\bibfnamefont{C.}~\bibnamefont{Berger}} \bibnamefont{et~al.},
  \bibinfo{journal}{Science} \textbf{\bibinfo{volume}{312}},
  \bibinfo{pages}{1191} (\bibinfo{year}{2006}).

\bibitem[{\citenamefont{Mallet et~al.}(2007)}]{mallet-2007}
\bibinfo{author}{\bibfnamefont{P.}~\bibnamefont{Mallet}} \bibnamefont{et~al.},
  \emph{\bibinfo{title}{{Electron states of mono- and bilayer graphene on SiC
  probed by STM}}} (\bibinfo{year}{2007}), \bibinfo{note}{cond-mat/0702406}.

\bibitem[{\citenamefont{Beenakker}(2006)}]{beenakker:067007}
\bibinfo{author}{\bibfnamefont{C.~W.~J.} \bibnamefont{Beenakker}},
  \bibinfo{journal}{Phys. Rev. Lett.} \textbf{\bibinfo{volume}{97}},
  \bibinfo{pages}{067007} (\bibinfo{year}{2006}).

\bibitem[{\citenamefont{Wehling and Balatsky}(2007)}]{technote07}
\bibinfo{author}{\bibfnamefont{T.~O.} \bibnamefont{Wehling}} \bibnamefont{and}
  \bibinfo{author}{\bibfnamefont{A.~V.} \bibnamefont{Balatsky}}
  (\bibinfo{year}{2007}), \bibinfo{note}{(unpublished)}.

\end{thebibliography}
\end{document}